\documentclass[12pt,journal,draftcls,a4paper,onecolumn]{IEEEtran} 
\usepackage{dsfont}
\usepackage{subeqnarray}
\usepackage{cleveref}
\usepackage{amssymb}
\usepackage{amsfonts}
\usepackage{amsmath}
\usepackage{lipsum}
\usepackage{graphicx}
\usepackage{amsmath}
\usepackage[all,poly]{xy}
\usepackage{multirow}
\usepackage{algorithm}
\usepackage{algorithmic}
\usepackage{color}
\usepackage[noadjust]{cite}

\title{Bayesian nonlinear hyperspectral unmixing with
spatial residual component analysis
}
\author{Yoann Altmann, Marcelo Pereyra and Stephen McLaughlin
\thanks{This study was supported by the Direction G\'en\'erale de l'armement, French Ministry of Defence, by the SuSTaIN program - EPSRC grant EP/D063485/1 - at the Department of Mathematics, University of Bristol, and the EPSRC via grant EP/J015180/1.}
\thanks{Y. Altmann and S. McLaughlin are with the School of Engineering and Physical Sciences, Heriot-Watt University, Edinburgh
U.K. (email: \{Y.Altmann,S.McLaughlin\}@hw.ac.uk).}
\thanks{M. Pereyra is with the School of Mathematics of the University of Bristol, Bristol
U.K. (email: marcelo.pereyra@bristol.ac.uk).}}

\newcommand{\red}{\textcolor{red} }
\newcommand{\blue}{\textcolor{blue} }
\newcommand{\bmu}{\boldsymbol{\mu}}

\newcommand{\bbeta}{\boldsymbol{\beta}}
\newcommand{\bgam}{\boldsymbol{\gamma}}
\newcommand{\bGam}{\boldsymbol{\Gamma}}

\newcommand{\bfsigma}{\boldsymbol{\sigma}}
\newcommand{\bSigma}{\boldsymbol{\Sigma}}
\newcommand{\bPsi}{{\boldsymbol \Psi}}

\newcommand{\bphi}{{\boldsymbol \phi}}

\def\bfx{{\mathbf{x}}}

\def\bfG{{\mathbf{G}}}

\def\bfR{{\mathbf{R}}}
\def\bfS{{\mathbf{S}}}

\def\bfW{{\mathbf{W}}}

\def\bbR{{\mathbb{R}}}

\newcommand{\Vpix}[1]{\mathbf{y}_{#1}}

\newcommand{\MATpix}{\mathbf{Y}}

\newcommand{\Vpixels}{\mathbf{y}}

\newcommand{\nbband}{L}

\newcommand{\nbmat}{R}
\newcommand{\nomat}{r}

\newcommand{\MATmat}{{\mathbf M}}
\newcommand{\Vmat}[1]{{\mathbf m}_{#1}}
\newcommand{\mat}[2]{m_{#1,#2}}

\newcommand{\MATabond}{{\mathbf A}}

\newcommand{\abond}[2]{{a}_{#1,#2}}
\newcommand{\Vabond}[1]{{\boldsymbol{a}}_{#1}}

\newcommand{\Vnoise}{{\mathbf e}}

\newcommand{\paramvect}{\boldsymbol{\theta}}

\newcommand{\transp}{^T}

\newcommand{\diag}[1]{\textrm{diag}\left(#1\right)}
\newcommand{\Ndistr}[1]{\mathcal{N}\left(#1\right)}

\newcommand{\norm}[1]{\left\|#1\right\|}

\newcommand{\Vzero}{\boldsymbol{0}}
\newcommand{\Id}[1]{\textbf{I}_{#1}}
\newcommand{\Indicfun}[2]{\textbf{1}_{#1}\left(#2\right)}

\newenvironment{algogo}[1]{
\smallskip
\noindent \hrule\vspace{0.2\baselineskip} \hrule
\begin{small}
\refstepcounter{algo} \center{\bf \textsc{Algorithm \thealgo}}
\\{\center{\bf #1}}
\smallskip
\flushleft
 } {
\end{small}
\smallskip
\hrule\vspace{0.2\baselineskip} \hrule
\smallskip
}

\newcounter{algo}
\renewcommand{\thealgo}{\arabic{algo}}

\begin{document}
\maketitle

\begin{abstract}
This paper presents a new Bayesian model and algorithm for nonlinear unmixing of hyperspectral images. The model proposed represents the pixel reflectances 
as linear combinations of the endmembers, corrupted by nonlinear (with respect to the endmembers) terms and additive Gaussian noise. Prior knowledge about the problem is embedded in a hierarchical model that describes the dependence structure between the model parameters and their constraints. In particular, a gamma Markov random field is used to model the joint distribution of the nonlinear terms, which are expected to exhibit significant spatial correlations. An adaptive Markov chain Monte Carlo algorithm is then proposed to compute the Bayesian estimates of interest and perform Bayesian inference. This algorithm is equipped with a stochastic optimisation adaptation mechanism that automatically adjusts the parameters of the gamma Markov random field by maximum marginal likelihood estimation. Finally, the proposed methodology is demonstrated through a series of experiments with comparisons using synthetic and real data and  with competing state-of-the-art approaches.
\end{abstract}

\begin{keywords}
Hyperspectral imagery, nonlinear spectral unmixing, residual
component analysis, Gamma Markov random field, Bayesian estimation.
\end{keywords}

\section{Introduction}
Spectral unmixing (SU) is a key problem in the analysis of hyperspectral images. This is a source separation problem consisting of recovering 
the spectral signatures (endmembers) of the materials present in the scene, and quantifying their proportions within each hyperspectral image pixel. The SU problem has been widely
studied for images where pixel reflectances are linear combinations of pure component spectra \cite{Heinz2001,Bioucas2012}. However, it is now widely accepted that the linear 
mixing model (LMM) can be inappropriate for some hyperspectral images, particularly those containing sand-like materials or relief. Several nonlinear mixing models (NLMM) have been recently proposed to address the limitations of the LMM. There are two main approaches to dealing with NLMM. The first seeks to model the physics of the image formation model (e.g.,  \emph{intimate mixtures} \cite{Hapke1981} for short-range multiple light scattering, and polynomial models for long-range multiple light scattering \cite{Somers2009,Nascimento2009,Halimi2010,Meganem2013}). The second seeks to construct flexible models that can represent a wide range of nonlinearities. This is can be achieved using neural networks, kernel functions \cite{Chen2012}, or post-nonlinear transformations \cite{Altmann2012a,Altmann2014b} for instance.

While the consideration of nonlinear effects can be very relevant in some specific regions of the scene, most hyperspectral image pixels are well described by the LMM. Therefore, models for nonlinear unmixing should include the LMM as a special case. Here we use a variation of the Bayesian NLMM proposed recently in Altmann et al. \cite{Altmann2014a}, which is inspired by residual component analysis (RCA) \cite{Kalaitzis2012}. In that model the nonlinear effects in hyperspectral images are represented as additive perturbations (of the LMM) that are modelled as a collection of Gaussian processes (GPs) combined with a hidden Potts-Markov random field (MRF) partitioning the image into regions sharing the same GP. The model of \cite{Altmann2014a} has two drawbacks that we address in this paper. First, the Potts model leads to a piecewise constant representation that constrains nonlinearities to take a finite number of possible energy states (the so-called \emph{nonlinearity levels}); this number is difficult to specify a priori unless there is very accurate knowledge about the nonlinearities present in the scene. Second, in \cite{Altmann2014a} nonlinearities are allowed to take negative values, as this allows marginalising them analytically (i.e., integrating them out of the model) and thus simplifies the statistical inference procedure;. However, our experiments suggest that taking into account the assumption that the nonlinearities are positive can improve the estimation results significantly when the nonlinear terms are positive (see \cite{Meganem2013} for more details about this positivity assumption). Here we address these drawbacks by replacing the Potts MRF by a gamma MRF model \cite{Dikmen2010}. This model has the key advantages of 1) promoting spatial regularity in the nonlinearity terms without enforcing a piecewise constant representation with a finite number of levels, and 2) it can easily incorporate a positivity constraint for the nonlinearities.

The remainder of the paper is organised as follows. Section \ref{sec:Problem} recalls the RCA model for hyperspectral image unmixing. The Bayesian NLMM proposed in this paper is presented in Section \ref{sec:bayesian}. In Section \ref{sec:Gibbs} we propose a Markov chain Monte Carlo Bayesian algorithm to perform statistical inference in this model and 
we define Bayesian estimators for nonlinear unmixing and nonlinearity detection. Sections \ref{sec:simu_synth} and \ref{sec:simu_real} demonstrate the proposed methodology through a series of experiments with synthetic and real hyperspectral images and comparisons with methods from the state of the art. Conclusions and perspectives for future work are finally reported in Section \ref{sec:conclusion}.

\section{Problem formulation}
\label{sec:Problem} 
Let $\Vpix{i,j} \in \bbR^{\nbband}$ be the pixel at location $(i,j)$ of an hyperspectral image $\MATpix$ of size $N_{\textrm{row}} \times N_{\textrm{col}}$ and observed at $L$ spectral bands. We model each image pixel as a linear combination of $R$ known spectra or endmembers $\Vmat{r}$, plus an additive perturbation $\bphi_{i,j}$ embedding nonlinearities and additive noise
\begin{eqnarray}
\label{eq:NLM0}
\Vpix{i,j} & = & \sum_{r=1}^{R} \abond{r}{i,j}\Vmat{r} + \bphi_{i,j}+ \Vnoise_{i,j}\nonumber\\
 & = &  \MATmat \Vabond{i,j} + \bphi_{i,j}+ \Vnoise_{i,j}, \quad \forall (i,j)
\end{eqnarray}
where $\Vmat{\nomat} =
[\mat{\nomat}{1},\ldots,\mat{\nomat}{\nbband}]\transp$ is the
spectral response of the $\nomat$th material present in the scene, $\abond{r}{i,j}$ is its abundance within pixel $(i,j)$ and $\Vnoise_n \sim
\Ndistr{\Vzero_{\nbband},\bSigma_0}$ is Gaussian noise with diagonal covariance matrix $\bSigma_0=\diag{\bfsigma^2}$ with elements $\bfsigma^2=[\sigma_1^2,\ldots,\sigma_L^2]\transp$ (note that matrix and vector notations $\MATmat =
[\Vmat{1},\ldots,\Vmat{\nbmat}]$ and
$\Vabond{i,j}=[\abond{1}{i,j},\ldots, \abond{\nbmat}{i,j}]\transp$ have
been used in the second row of \eqref{eq:NLM0}). Due to physical considerations we model the abundances as non-negative quantities and set $\abond{r}{i,j} \in \mathbb{R}^+$ (notice that because we consider non-linear mixing we do not use the sum-to-one constraint that is commonly enforced in linear mixing models). Moreover, for the nonlinear effects we use the deterministic model
\begin{eqnarray} 
\label{eq:nonlin}
\bphi_{i,j} = \bphi(\bgam_{i,j}) & = & \sum_{k=1}^{R-1} \sum_{k'=k+1}^{R}\gamma_{i,j}^{(k,k')}\sqrt{2}\Vmat{k}\odot\Vmat{k'} \nonumber\\
& + & \sum_{k=1}^R \gamma_{i,j}^{(k)}\Vmat{k}\odot\Vmat{k}.
\end{eqnarray}
that is parametrised by a vector of nonlinearity coefficients $\bgam_{i,j}=[\gamma_{i,j}^{(1,2)}, \ldots, \gamma_{i,j}^{(R-1,R)},  \gamma_{i,j}^{(1)},\ldots,\gamma_{i,j}^{(R)}]\transp$ of length $K = R(R+1)/2$. This choice of model is motivated by the fact that the nonlinearities in hyperspectral images are well modelled by polynomial interactions between endmembers, which provides a flexible representation that can approximate a wide range of nonlinear effects (see \cite{Nascimento2009,Fan2009,Halimi2010,Altmann2012a,Meganem2013} for more details). Moreover, in this paper we assume that $\gamma_{i,j} \in \mathbb{R}^+$ because of the considerations reported in \cite{Meganem2013} and because we have observed that it can improve estimation results significantly. However, in Section V we also describe a version of our model where this positivity constraint in relaxed. Also note that the factors $\sqrt{2}$ in \eqref{eq:nonlin} are simply introduced to simplify kernel computations \cite{Altmann2012a}, however these factors do not have a physical interpretation and can be removed from \eqref{eq:nonlin} without changing the model by scaling the coefficients $\gamma_{i,j}^{(k,k')}$ appropriately.

This paper considers the inverse problems of estimating the abundances $\Vabond{i,j}$ and of detecting the presence of nonlinearities at each image pixel $\Vpix{i,j}$ 
(whose intensity can then be measured by estimating $\| \bgam_{i,j} \|^2_2$). We formulate this problem as a statistical inference task that we address in a Bayesian framework by defining an appropriate Bayesian model and inference algorithm.

\section{Bayesian model}
\label{sec:bayesian}
This section presents an original Bayesian model for inferring the unknown quantities of interest $\MATabond$ and $\bGam$  from the observed hyper-spectral image $\MATpix$, where  $\MATabond$ is an $R \times N_{\textrm{row}} \times N_{\textrm{col}}$ array gathering the abundance vectors $\Vabond{i,j}$ and $\bGam$ an $K \times N_{\textrm{row}} \times N_{\textrm{col}}$ array gathering the nonlinearity coefficient vectors $\bgam_{i,j}$. Following a hierarchical Bayesian approach, we also include in the model all the parameters of the model whose values are not easily known a priori and need to be inferred from data jointly with $\MATabond$ and $\bGam$ (e.g., the noise covariance $\bfsigma^2$). Unlike $\MATabond$ and $\bGam$, the other unknown quantities are of no interest for decision making and are therefore removed from the model by marginalisation during the inference procedure. 
\subsection{Likelihood}
From the non-linear mixing model \eqref{eq:NLM0}, and by assuming that observations $\MATpix=[\Vpix{1},\ldots,\Vpix{N}]$  are conditionally independent given $\MATabond, \bGam$ and $\bfsigma^2$, we obtain
\begin{eqnarray}\label{eq:likelihood}
  f(\MATpix|\MATabond, \bGam, \bfsigma^2) \propto \prod_{i,j}|\bSigma_0|^{-1/2}\exp\left[-\dfrac{(\Vpix{i,j} - \boldsymbol{x}_{i,j})\transp\bSigma_0^{-1}(\Vpix{i,j}- \boldsymbol{x}_{i,j})}{2}\right]
\end{eqnarray}
with $\boldsymbol{x}_{i,j} = \MATmat \Vabond{i,j} + \bphi(\gamma_{i,j})$, $\bSigma_0=\diag{\bfsigma^2}$, and where $\propto$ denotes proportionality. Note that to lighten notation the dependence on $\MATmat$ is not denoted explicitly ($\MATmat$ is assumed to be perfectly known).

\subsection{Prior for the abundance matrix $\MATabond$}
We assign the abundance coefficients the following hierarchical prior distribution
\begin{eqnarray}
\label{eq:abond_prior}
\abond{r}{i,j} | \beta_{r} &\sim &\mathcal{N}_{\bbR^+}(0,\beta_{r})\\
\beta_{r} &\sim &\mathcal{IG}(\alpha_1,\alpha_2)
\end{eqnarray}
parametrised by some fixed hyper-parameters $\alpha_1$ and $\alpha_2$, and where $\mathcal{N}_{\bbR^+}(0,\beta_{r})$ denotes the truncated Gaussian distribution on $\bbR^+$ with mode $0$ and scale parameter $\beta^{1/2}_{r}$, reflecting the positivity of $\abond{r}{i,j}$. This prior model is very flexible and can be adjusted to represent a wide variety of prior beliefs. Without loss of generality, here we set $\alpha_1 = 1$ and $\alpha_2 = 2$, leading to a (marginal) exponential prior for $\abond{r}{i,j}$ that represents our prior beliefs that abundances are proportions and take values mainly in $[0,1]$. In particular that we expect small values to occur more frequenty because most materials are not present on all image pixels (notice however that the exact values of $\alpha_1$ and $\alpha_2$ generally have little impact on the inference because $\MATabond$ is very high dimensional and dominates the distribution of $\beta_{r}$).

A strength of hierarchical priors such as \eqref{eq:abond_prior} is their natural capacity to encode prior dependences between unknown variables. For example, we expect the abundance coefficients associated with the same material to exhibit correlations, particularly in terms of their scale. This belief is encoded in  \eqref{eq:abond_prior} by defining one common hidden variable $\beta_r$ for each material or endmember $\Vmat{r}$, which is shared by all the abundances related to that material. This hierarchical structure operates as a global pooling mechanism that shares information across the rows of $\MATabond$ (i.e, the abundance coefficients associated to the $r$th material) to improve estimation performance. It is also possible to relate \eqref{eq:abond_prior} to a group $\ell_1$ regularisation or a composite $\ell_1-\ell_2$ regularisation, in the sense that without the pooling mechanism marginalising the hidden variables $\beta_{r}$ would lead to an $\ell_1$ regularisation for the abundances $\abond{r}{i,j}$, and introducing the pooling mechanism also links the rows of $\MATabond$ at the level of their $\ell_2$ norms. Finally, it is also worth mentioning that model \eqref{eq:abond_prior} does not account explicitly for spatial correlations between the abundance vectors. This information could be introduced into the model by using mixtures of Gaussian or Dirichlet distributions \cite{Eches2011,Nascimento2012}. However, the main focus of this paper is the consideration of the spatial dependence between the nonlinearities and their impact on estimation performance, though we hope and anticipate that future models will exploit both types of spatial information.

Finally, assuming that abundances are prior independent given the hidden variables $\bbeta=[\beta_1,\ldots,\beta_R]\transp$, we obtain the following joint prior for $\MATabond, \bbeta$ 
\begin{eqnarray}
\label{eq:joint_abund_prior}
f(\MATabond,\bbeta) = f(\MATabond|\bbeta)f(\bbeta)
\end{eqnarray}
with $f(\MATabond|\bbeta)  = \prod_{r,i,j}f(\abond{r}{i,j}|\beta_r)$ and $f(\bbeta) = \prod_{r} f(\beta_r|\alpha_1,\alpha_2)$. Also notice that by using the hierarchical structure \eqref{eq:abond_prior} we obtain conjugate priors and hyper-priors for $\abond{r}{i,j}$ and $\beta_r$. Conjugacy generally leads to inference algorithms with significantly better tractability and computational efficiency, which is crucial given the high dimensionality of $\MATabond$.

\subsection{Priors for the nonlinearity coefficients $\bGam$}
One of the contributions of this paper is to propose the following hierarchical prior for the nonlinearity coefficients
\begin{eqnarray}
\label{eq:prior2_gam_preliminary}
\left\{
\begin{array}{l}
\bgam_{i,j} | s_{i,j} \sim  \mathcal{N}_{\bbR_+^K}\left(\Vzero, s_{i,j} \Id{K}\right)\\
s_{i,j} \sim  \mathcal{IG}(\alpha_3,\alpha_3 \alpha_{4,i,j})
\end{array}
\right.
\end{eqnarray}
where $\bbR_+^K$ denotes the $K$-dimensional positive orthant, reflecting a positivity constraint on $\bgam_{i,j}$. Notice that this prior is parametrised by a local hyper-parameter $\alpha_{4,i,j}$ that is related to the prior mean of $s_{i,j}$, and therefore to the average power of the nonlinearities at the pixel $(i,j)$ (via the norm $\|\bgam_{i,j}\|_2^2$). The prior also depends on a global hyper-parameter $\alpha_3$ that controls the shape of the tails of the prior \eqref{eq:prior2_gam_preliminary}, and therefore the probability of large deviations between $s_{i,j}$ and $\alpha_{4,i,j}$. A careful selection of $\alpha_{4,i,j}$ and $\alpha_3$ will allow exploiting the spatial dependences between the nonlinearity coefficients $\bgam_{i,j}$ to improve estimation performance.

\begin{figure}[ht]
\centering
\includegraphics[width=\columnwidth]{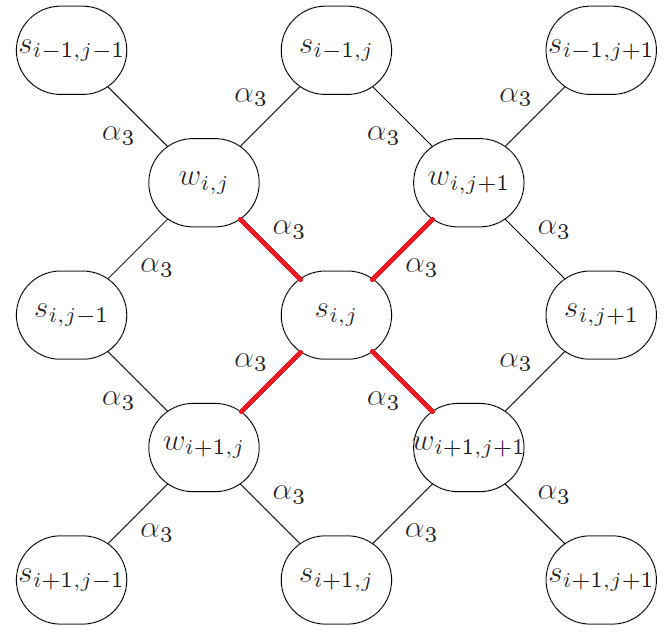}
\caption{Proposed $1$st order neighbourhood structure ($\forall (i,j) \in \Omega$).}
\label{fig:neighbour}
\end{figure}

As explained previously, a key feature of hierarchical models is their capacity to encode dependences and act as pooling mechanisms that share information across covariates to improve the inference. Here we wish to specify \eqref{eq:prior2_gam_preliminary} to reflect the prior belief that nonlinearities exhibit spatial correlations. In particular, due to the spatial organisation of images, we expect the values of $\bgam_{i,j} $ to vary smoothly from one pixel to another and exhibit occasional abrupt and sharp changes. In order to model this behaviour we specify $\alpha_{4,i,j}$ such that the resulting prior for $\bGam$ is a hidden gamma-Markov random field (GMRF) \cite{Dikmen2010}. More precisely, we denote by $\bfS$ the $N_{\textrm{row}} \times N_{\textrm{col}}$ matrix with elements $s_{i,j}$, introduce a $(N_{\textrm{row}}+1) \times (N_{\textrm{col}}+1)$ auxiliary matrix $\bfW$ with elements $w_{i,j} \in \mathbb{R}^+$ and define a bipartite conditional independence graph between $\bfS$ and $\bfW$ such that each $s_{i,j}$ is connected to four neighbour elements of $\bfW$ and vice-versa. This $1$st order neighbourhood structure is depicted in Fig.\ref{fig:neighbour}, where we notice that any given $s_{i,j}$ and $s_{i+1,j}$ are $2$nd order neighbours via $w_{i,j+1}$ and $w_{i+1,j+1}$. The role of these auxiliary variables is to introduce positive dependence between the neighbouring elements of $s_{i,j}$ and therefore to promote regularity. However, this model also allows occasional sharp changes because the distribution of any $s_{i,j}$ given its neighbours in $s_{i+1,j}$ is heavy-tailed.  A GMRF prior for $\bfS,\bfW$ \cite{Dikmen2010} is then defined as the following hierarchical prior \cite{Dikmen2010}:
\begin{subeqnarray}
\label{eq:prior1_gam}
\slabel{eq:prior1_gam1}
\bgam_{i,j} | s_{i,j} &\sim & \mathcal{N}_{\bbR_+^K}\left(\Vzero, s_{i,j} \Id{K}\right)\\
\slabel{eq:prior1_gam2}
s_{i,j}|\bfW,\alpha_3 &\sim & \mathcal{IG}(\alpha_3,\alpha_3 \alpha_{4,i,j}(\bfW))\\
\slabel{eq:prior1_gam3}
w_{i,j}|\bfS,\alpha_3 &\sim & \mathcal{G}(\alpha_3, 1/(\alpha_3\alpha_{5,i,j}(\bfS)))
\end{subeqnarray}
where
\begin{eqnarray*}
\alpha_{4,i,j}(\bfW) &=& w_{i,j} + w_{i+1,j} + w_{i,j+1} + w_{i+1,j+1}/4\\
\alpha_{5,i,j}(\bfW) &=& (s_{i,j}^{-1} + s_{i-1,j}^{-1} + s_{i,j-1}^{-1} + s_{i-1,j-1}^{-1})/4.
\end{eqnarray*}
The density for this joint prior for $\bGam$, $\bfS$ and $\bfW$ is given by
\begin{eqnarray}
f(\bGam,\bfS,\bfW | \alpha_3) = f(\bGam|\bfS)f(\bfS,\bfW|\alpha_3)\nonumber
\end{eqnarray}
where $f(\bGam|\bfS) = \prod_{i,j} f(\bgam_{i,j}|s_{i,j})$ and 
\begin{eqnarray}
\label{eq:GMRF}
f(\bfS,\bfW|\alpha_3) & = & \dfrac{1}{Z(\alpha_3)} \prod_{(i,j) \in \mathcal{V}_{\bfS}} 
\left(s_{i,j}\right)^{-\left(\alpha_3+1 \right)} \nonumber\\
& \times & \prod_{(i',j') 
\in \mathcal{V}_{\bfW}} w_{i',j'}^{\left(\alpha_3-1 \right)}\nonumber\\
 & \times & \prod_{\left((i,j),(i',j')\right) \in \mathcal{E}} \exp 
\left(\dfrac{-\alpha_3 w_{i',j'}}{4 s_{i,j}} \right).
\end{eqnarray}
Notice that we denote explicitly the dependence on the value of $\alpha_3$, which here acts a regularisation parameter that controls the amount of spatial smoothness enforced by the GMRF. Following an empirical Bayesian approach, the value of $\alpha_3$ remains unspecified and will be adjusted automatically during the inference procedure by maximum marginal likelihood estimation using the technique \cite{Pereyra2014ssp}. We refer to this model as \emph{gamma-RCA} with positivity constraint (G-RCA+).

Finally, it is worth mentioning that this model for the nonlinearity coefficients has similarities with the model proposed in \cite{Altmann2014a} that also considers the spatial regularity of non-linearities. However, the model described \cite{Altmann2014a} follows a segmentation approach in which the non-linearity coefficients are assumed to (and constrained) to take values in a finite set of possible values. This leads to a piece-wise constant representation and requires specification of the number of nonlinearity levels present in the image, a value that is often difficult to determine a priori. The model proposed in this paper provides a spatially smooth representation of the nonlinearities that is possibly more realistic than the piece-wise constant representation of  \cite{Altmann2014a}, and also has the practical advantage of not requiring practitioners to specify the finite number of admissible nonlinearity levels. Another important distinction is that the model described in \cite{Altmann2014a} does not allow the non-negativity constraint for $\bgam_{i,j}$ to be introduced, which we have found to improve significantly the estimation of the nonlinearities when the nonlinear coefficients (see \cite{Meganem2013} for more details about this positivity assumption).

For potential applications where the assumption of positive nonlinear terms would not hold, the proposed model G-RCA+ can be modified to allow $\bgam_{i,j}$ to take positive and negative values in $\mathbb{R}$ by using the following hierarchical prior
\begin{subeqnarray}
\label{eq:prior2_gam}
\slabel{eq:prior2_gam1}
\bgam_{i,j} | s_{i,j} &\sim & \mathcal{N}\left(\Vzero, s_{i,j} \Id{K}\right)\\
\slabel{eq:prior2_gam2}
s_{i,j}|\bfW,\alpha_3 &\sim & \mathcal{IG}(\alpha_3,\alpha_3 \alpha_{4,i,j}(\bfW))\\
\slabel{eq:prior2_gam3}
w_{i,j}|\bfS,\alpha_3 &\sim & \mathcal{G}(\alpha_3, 1/(\alpha_3\alpha_{5,i,j}(\bfS))).
\end{subeqnarray}
Notice that now $\bgam_{i,j} | s_{i,j}$ is Gaussian, instead of truncated Gaussian as in \eqref{eq:prior1_gam1}. This modification leads to a gamma-RCA model without positive constraint (G-RCA) that bridges between the proposed G-RCA+ model and the original RCA model \cite{Altmann2014a}, where nonlinearity coefficients are constrained to take a finite number of positive and negative values. For brevity, we henceforth consider the G-RCA+ model unless stated otherwise.

\subsection{Prior for the noise covariance $\bfsigma^2$}
We assume that there is no prior knowledge available about the values of noise covariance (other than the fact that it is diagonal) and assign each diagonal element $\sigma_{\ell}^2$ a Jeffreys' prior, leading to the joint prior
\begin{eqnarray}
f(\bfsigma^2) = \prod_{\ell=1}^L f(\sigma_{\ell}^2), \quad \textrm{with} \quad f(\sigma_{\ell}^2) \propto \sigma_{\ell}^{-2} \Indicfun{\bbR^+}{\sigma_{\ell}^2}.
\end{eqnarray}
In scenarios where prior knowledge is available, practitioners can incorporate this information into the model by replacing the Jeffreys' prior by a more informative model (e.g. a conjugate inverse-Gamma distribution).

\subsection{Posterior distribution}
\begin{figure}[!ht]
\centerline{ \xymatrix{
 *+<0.05in>+[F-]+{\alpha_1} \ar@/^/[rd] &  *+<0.05in>+[F-]+{\alpha_2} \ar@/^/[d] & \alpha_3 \ar@/^/[d]&  \\
  & \bbeta \ar@/^/[d]   & (\bfS,\bfW) \ar@/^/[d] & \\
  *+<0.05in>+[F-]+{\MATmat} \ar@/_/[rd]    & \MATabond \ar@/_/[d] &   \bGam \ar@/^/[ld]& \bfsigma^2 \ar@/^/[lld]  \\
   & \MATpix &   & }
} \caption{Graphical model for the proposed hierarchical Bayesian model (fixed quantities appear in boxes).} \label{fig:DAG}
\end{figure}
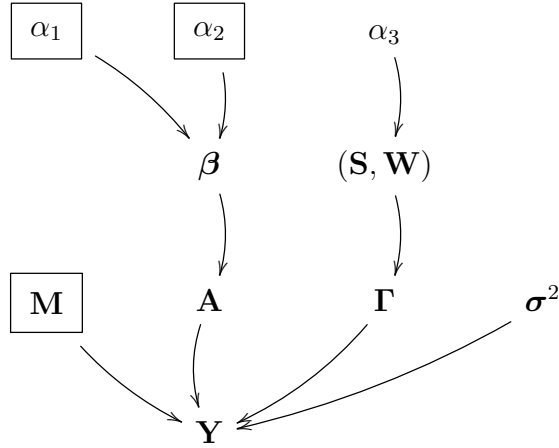

We are now ready to specify the posterior distribution for $\MATabond, \bGam, \bfsigma^2, \bfS,\bfW$ and $\bbeta$ given the observed hyper-spectral image $\MATpix$ and the value of the spatial regularisation hyper-parameter $\alpha_3$ (recall that this value will be determined by maximum marginal likelihood estimation during the inference procedure). Using Bayes' theorem, and assuming prior independence between $(\MATabond,\bbeta)$, $(\bGam,\bfS,\bfW)$ and $\bfsigma^2$, the joint posterior distribution associated with the proposed Bayesian model is given by
\begin{eqnarray}
\label{eq:posterior}
f(\MATabond, \bGam,\bfsigma^2, \bfS,\bfW, \bbeta|\MATpix,\alpha_3) &\propto f(\MATpix|\MATabond, \bGam, \bfsigma^2)f(\MATabond|\bbeta)f(\bbeta)f(\bGam|\bfS)f(\bfS,\bfW|\alpha_3).
\end{eqnarray}
For illustration, Fig. \ref{fig:DAG} depicts the directed acyclic graph (DAG) summarising the structure proposed Bayesian model (recall that $\bfS,\bfW$ have a bi-partite neighbourhood structure, which is illustrated in the graphical model of Fig. \ref{fig:neighbour}).


\section{Bayesian Inference}\label{sec:Gibbs}
\subsection{Bayesian estimators}
\label{subsec:Bayesian_estimators}
The Bayesian model defined in Section \ref{sec:bayesian} specifies the joint posterior density for the unknown parameters $\MATabond, \bGam,\bfsigma^2, \bfS,\bfW$ and $\bbeta$ given the observed quantities $\MATpix, \MATmat$ and the hyper-parameter $\alpha_3$. This posterior distribution models our complete knowledge about the unknowns given the observed data and the prior information available. In this section we define suitable Bayesian estimators to summarise this knowledge and perform hyperspectral unmixing. More precisely, we propose the following two Bayesian estimators for hyperspectral non-linear unmixing and nonlinearity estimation and detection: The marginal posterior mean or minimum mean square error estimator of the abundance matrix
\begin{eqnarray}\label{abundanceMMSE}
\hat{\MATabond}_{\textrm{MMSE}} =  \textrm{E}\left[\MATabond | \MATpix,\hat{\alpha}_3\right],
\end{eqnarray}
where the expectation is taken with respect to the marginal posterior density $f(\MATabond|\MATpix,\alpha_3)$ (by marginalising $\bGam,\bfsigma^2, \bfS,\bfW$ and $\bbeta$ this density takes into account their uncertainty). The minimum mean square error estimator of the pixel-wise nonlinearity energy 
\begin{eqnarray}\label{phiMMSE}
\left(\widehat{\norm{\bphi_{i,j}}}_2^2\right)_{\textrm{MMSE}} = \textrm{E}\left[\norm{\bphi_{i,j}}_2^2| \MATpix,\hat{\alpha}_3\right],
\end{eqnarray}
where the expectation is now taken with respect to the marginal posterior density $f(\norm{\bphi_{i,j}}^2 |\MATpix,\alpha_3)$. And the Bayesian hypothesis test for nonlinearity detection
$$
P_{i,j} > a_1/(a_0 + a_1),
$$
where
\begin{eqnarray}\label{eq:nonlin_proba}
P_{i,j}=\textrm{E}\left[ T_{i,j}(\bphi_{i,j},\Vabond{i,j}) > \eta \, | \MATpix,\hat{\alpha}_3 \right],
\end{eqnarray}
with
\begin{eqnarray*}
T_{i,j}(\bphi_{i,j},\Vabond{i,j}) = \dfrac{\norm{\bphi_{i,j}}_2^2}{\norm{\Vpix{i,j}-\MATmat\Vabond{i,j}-\bphi_{i,j}^{(t)}}_2^2},
\end{eqnarray*} 
is the posterior probability that the power of the nonlinear effects in pixel $(i,j)$ is $\eta$ times larger than the power of the noise at that pixel, and where $a_0$ and $a_1$ are application-specific weights associated with incorrectly rejecting or accepting this hypothesis (in our experiments we have used $a_0 = a_1, \eta = 2$).

Notice that in \eqref{abundanceMMSE}, \eqref{phiMMSE} and \eqref{eq:nonlin_proba} we have set $\alpha_3 = \hat{\alpha_3}$, which denotes the maximum marginal likelihood estimator of the MRF regularisation hyper-parameter $\alpha_3$ given the observed data $\MATpix$, i.e.,
\begin{eqnarray}\label{alphaML}
\hat{\alpha}_3=\underset{\alpha_3 \in \mathbb{R}^+}{\textrm{argmax}} f\left(\MATpix | \alpha_3\right),
\end{eqnarray}
This approach for specifying $\alpha_3$ is taken from the empirical Bayes framework in which hyper-parameters with unknown values are replaced by point estimates computed from observed data (as opposed to being fixed a priori or integrated out of the model by marginalisation). As explained in \cite{Pereyra2014ssp}, this strategy has several important advantages for MRF hyper-parameters with doubly intractable conditional distributions such as $\alpha_3$. In particular, it allows for the automatic adjustment of the value of $\alpha_3$ for each image (thus producing significantly better estimation results than using a single fixed value of $\alpha_3$ for all images), and has a computational cost that is several times lower than that of competing approaches, such as including $\alpha_3$ in the model and subsequently marginalising it during the inference procedure \cite{Pereyra2013ip}.

\subsection{Bayesian algorithm}
Computing the estimators \eqref{abundanceMMSE}, \eqref{phiMMSE} and \eqref{eq:nonlin_proba} is very challenging because it involves calculating expectations with respect to posterior marginal densities, which in turn require evaluating the full posterior \eqref{eq:posterior} and integrating it over a very high-dimensional space. Computing $\hat{\alpha}_3$ is also difficult because it involves solving an intractable optimisation problem, (because it is not possible to evaluate the marginal likelihood $f(\MATpix | \alpha_3)$ or its gradient $\nabla f(\MATpix | \alpha_3)$). Here we adopt the approach proposed in \cite{Pereyra2014ssp} and design a stochastic optimisation and simulation algorithm to compute \eqref{abundanceMMSE}, \eqref{phiMMSE}, \eqref{eq:nonlin_proba} and \eqref{eq:posterior} simultaneously. That is, we construct a stochastic gradient Markov chain Monte Carlo (SGMCMC) algorithm that simultaneously estimates $\hat{\alpha}_3$ and generates a chain of $N_{\textrm{MC}}$ samples $\{\MATabond^{(t)},\bGam^{(t)},\bfS^{(t)}\}_{t=1}^{N_{\textrm{MC}}}$ asymptotically distributed according to the marginal density $f(\MATabond, \bGam, \bfS | \MATpix,\hat{\alpha}_3)$ (this algorithm is summarised in Algo. \ref{algo:algo1} below). Once the samples have been generated, the estimators \eqref{abundanceMMSE}, \eqref{phiMMSE} and \eqref{eq:nonlin_proba} are approximated by Monte Carlo integration \cite[Chap. 10]{Robert2004}, i.e.,
\begin{eqnarray}
\label{eq:abundances_MC}
\hat{\MATabond}_{MMSEj} = \dfrac{1}{N_{\textrm{MC}}-N_{\textrm{bi}}}\sum_{t=N_{\textrm{bi}}+1}^{N_{\textrm{MC}}}
	\MATabond^{(t)},
\end{eqnarray}
\begin{eqnarray}
\label{eq:phi_MC}
\left(\widehat{\norm{\bphi_{i,j}}}_2^2\right)_{MMSEj} = \dfrac{1}{N_{\textrm{MC}}-N_{\textrm{bi}}}\sum_{t=N_{\textrm{bi}}+1}^{N_{\textrm{MC}}}
	\norm{\bphi\left(\bgam_{i,j}^{(t)}\right)}_2^2,
\end{eqnarray}
and
\begin{eqnarray}
\label{eq:estim_nonlin_proba}
\hat{P}_{i,j}=\dfrac{1}{N_{\textrm{MC}}-N_{\textrm{bi}}}\sum_{t=N_{\textrm{bi}}+1}^{N_{\textrm{MC}}}
	\left[\Indicfun{(\eta,\infty)}{T_{i,j}^{(t)}}\right],
\end{eqnarray}
with $T_{i,j}^{(t)}=\norm{\bphi\left(\bgam_{i,j}^{(t)}\right)}_2^2/\norm{\Vpix{i,j}-\MATmat\Vabond{i,j}^{(t)}-\bphi\left(\bgam_{i,j}^{(t)}\right)}_2^2$, and where the samples from the first $N_{\textrm{bi}}$ iterations (corresponding to the transient regime or burn-in period) are discarded. The main steps of this algorithm are detailed in below.\\

\subsubsection{Sampling the mixing parameters}
The conditional distribution of $\MATabond,\bGam$ given the other variables of the model is given by
\begin{eqnarray}
f(\MATabond,\bGam|\MATpix,\bPsi,\bfsigma^2,\alpha_3) = \prod_{i,j} f(\Vabond{i,j},\bgam_{i,j}|\Vpix{n},\bPsi,\bfsigma^2)
\end{eqnarray}
where
\begin{eqnarray}
\label{eq:post_A_GAM}
\Vabond{i,j},\bgam_{i,j}|\Vpix{n},\bPsi,\bfsigma^2 \sim \mathcal{N}_{\boldsymbol{\bbR_+^R+K}}(\bmu_{i,j},\bSigma_{i,j}),
\end{eqnarray}
with
\begin{eqnarray}
\left\{
    \begin{array}{lll}
	\bSigma_{i,j}& = & \left(\bfR_{i,j} + \bfG\transp\bSigma_0^{-1}\bfG \right)^{-1},\\
        \bmu_{i,j} & = & \bSigma_{i,j} \bfG\transp\bSigma_0^{-1} \Vpix{i,j},		
    \end{array}
\right.
\end{eqnarray}
and where $\bfG$ is an $L \times (R+D)$ matrix with the endmembers and the $D$ nonlinear interaction spectra, i.e., 
$\bfG=[\MATmat,\Vmat{1}\odot\Vmat{2},\ldots,\Vmat{R-1}\odot\Vmat{R},\Vmat{1}\odot\Vmat{1},\ldots,\Vmat{R}\odot\Vmat{R}]$ and 
\begin{eqnarray}
\bfR_{i,j} & = & \left( \begin{bmatrix}
  \textrm{diag}(\bbeta) & \Vzero_{R,D} \\
  \Vzero_{D,R} & s_{i,j}\Id{D}
 \end{bmatrix} \right)^{-1}.\nonumber
\end{eqnarray}
We simulate from \eqref{eq:post_A_GAM} using using the method proposed in \cite{Pakman2012}. Moreover, for the G-RCA model that does not constrain $\bgam_{i,j}$ to be positive, we replace  \eqref{eq:post_A_GAM} with the alternative conditional distribution for $\Vabond{i,j},\bgam_{i,j}|\Vpix{n},\bPsi,\bfsigma^2$ given by
\begin{eqnarray}
\label{eq:post_A_GAM_2}
\Vabond{i,j},\bgam_{i,j}|\Vpix{n},\bPsi,\bfsigma^2 \sim \mathcal{N}_{\boldsymbol{\bbR^R} \times \boldsymbol{\bbR^K}}(\bmu_{i,j},\bSigma_{i,j}),
\end{eqnarray}
that is also easy to sample using the method proposed in \cite{Pakman2012}.
\subsubsection{Sampling the noise variances}
The conditional distribution of $\bfsigma^2$ given the other variables of the model is given by
\begin{eqnarray}
\label{eq:post_bsigma2}
f(\bfsigma^2|\MATpix,\MATabond,\bGam,\bPsi,\alpha_3)  = \prod_{\ell=1}^{L} f(\sigma_{\ell}^2|\MATpix,\MATabond,\bPsi,\bbeta,\alpha_3),
\end{eqnarray}
with
\begin{eqnarray}
f(\sigma_{\ell}^2|\MATpix,\MATabond,\bGam,\bfS,\bfW,\bbeta,\alpha_3) = p_{\mathcal{IG}}\left(\sigma_{\ell}^2; N/2, \sum_{i,j} \dfrac{(\Vpix{i,j}-\bfx_{i,j})\transp\bSigma_0^{-1}(\Vpix{i,j}-\bfx_{i,j})}{2}\right).
\end{eqnarray}
Sampling from the conditional \eqref{eq:post_bsigma2} is achieved by simulating $L$ independent inverse gamma random variables.

\subsubsection{Sampling the abundance hyper-parameters}
Similarly, the elements of $\bbeta$ are also conditionally independent (given the other variables of the model) and can be simulated in parallel by generating inverse gamma random variables with distribution 
\begin{eqnarray}
\label{eq:post_beta}
\beta_{r}|\MATpix,\paramvect,\bfS,\bfW,a \sim \mathcal{IG}\left(\frac{N}{2}+\alpha_1, \sum_{i,j} \dfrac{\Vabond{i,j}^2}{2} + \alpha_2\right).
\end{eqnarray}

\subsubsection{Sampling the nonlinearity levels $\bfS$}
Again, the elements of $\bfS$ are conditionally independent given the other model parameters
\begin{eqnarray}
\label{eq:post_S}
f(\bfS|\MATpix,\paramvect,\bfW,\alpha_3) = \prod_{(i,j) \in \mathcal{V}_{\bfS}} f(s_{i,j}|\MATpix,\paramvect,\bfW,\alpha_3),
\end{eqnarray}
 and can be simulated in parallel by generating inverse gamma random variables with distribution 
\begin{eqnarray}
\label{eq:post_sij}
s_{i,j}|\MATpix,\paramvect,\bfW,\alpha_3 \sim
        \mathcal{IG}\left(\alpha_3 + \dfrac{K}{2}, \nu_{i,j} + \dfrac{\norm{\bgam_{i,j}}_2^2}{2} \right).
\end{eqnarray}

\begin{algogo}{Proposed MCMC algorithm}
     \label{algo:algo1}
     \begin{algorithmic}[1]
        \STATE \underline{Fixed input parameters:} Endmember matrix $\MATmat$, number of burn-in iterations $N_{\textrm{bi}}$, total number of iterations $N_{\textrm{MC}}$
				\STATE \underline{Initialization ($t=0$)}
        \begin{itemize}
        \item Set $\MATabond^{(0)},\bfsigma^{2(0)},\bGam^{(0)},\bfS^{(0)},\bfW^{(0)},\bbeta^{(0)},\alpha_3^{(0)}$
        \end{itemize}
        \STATE \underline{Iterations ($1 \leq t \leq N_{\textrm{MC}}$)}
        \STATE Sample $(\MATabond^{(t)},\bGam^{(t)})$ from \eqref{eq:post_A_GAM} if using G-RCA+, or\\
          Sample $(\MATabond^{(t)},\bGam^{(t)})$ from \eqref{eq:post_A_GAM_2} if using G-RCA
        \STATE Sample $\bfsigma^{2(t)}$ from \eqref{eq:post_bsigma2}
				\STATE Sample $\bbeta^{(t)}$ from \eqref{eq:post_beta}
				\STATE Sample $\bfS^{(t)}$ from \eqref{eq:post_S}
				\STATE Sample $\bfW^{(t)}$ from \eqref{eq:prior1_gam3}
				\IF{$t<N_{\textrm{bi}}$} 
				\STATE Sample $(\bfS',\bfW') \sim \mathcal{K}(\bfS,\bfW|\bfS^{(t)},\bfW^{(t)},\alpha_3^{(t-1)})$
				\STATE Set $\alpha_3^{(t)} = \mathcal{P}_{[0,A_t]}(\alpha_3^{(t-1)} + \delta_t \left[\Lambda(\bfS,\bfW)-\Lambda(\bfS',\bfW') \right])$\\
				with $\Lambda(\bfS,\bfW)= - \sum_{\left((i,j),(i',j')\right) \in \mathcal{E}} \dfrac{w_{i',j'}}{s_{i,j}} +  4 \left(\sum_{(i',j') 
\in \mathcal{V}_{\bfW}}\log\left(w_{i',j'}\right) - \sum_{(i,j) 
\in \mathcal{V}_{\bfS}}\log\left(s_{i,j}\right) \right)$
				\ELSE
				\STATE Set $\alpha_3^{(t)} = \alpha_3^{(t-1)}$
				\ENDIF
        \STATE Set $t = t+1$.
        \STATE Output $\{\MATabond^{(t)},\bGam^{(t)}, \bfS^{(t)} \}_{t=1}^{N_{\textrm{MC}}}$.
        \end{algorithmic}
\end{algogo}

\subsubsection{Updating the MRF regularisation parameter $\alpha_3$}
If marginal likelihood $f(\MATpix | \alpha_3)$ was tractable we could update $\alpha_3$ from one MCMC iteration to the next by using a classic gradient descent step
\begin{eqnarray*}
\alpha_3^{(t+1)} = \alpha_3^{(t)} + \delta_t \nabla \log f(\MATpix | \alpha_3^{(t)}),
\end{eqnarray*}
with $\delta_t = t^{-3/4}$, such that $\alpha_3^{(t)}$ converges to $\hat{\alpha}_3$ as $t \rightarrow \infty$. However, this gradient has two levels of intractability, one due to the marginalisation of $(\MATabond, \bGam,\bfsigma^2, \bfS,\bfW, \bbeta)$ and another one due to the intractable normalising constant of the gamma MRF. We address this difficulty by following the approach proposed in \cite{Pereyra2014ssp}; that is, by replacing $\nabla \log f(\MATpix | \alpha_3^{(t)})$ with an estimator computed with the samples generated by the MCMC algorithm at iteration $t$, and a set of two auxiliary variables  $(\bfS',\bfW') \sim \mathcal{K}(\bfS,\bfW|\bfS^{(t)},\bfW^{(t)},\alpha_3^{(t-1)})$ generated with an MCMC kernel $\mathcal{K}$ with target density \eqref{eq:GMRF} (in our experiments we used a Gibbs sampler implemented using a colouring scheme such that all the elements of $\bfS'$ and $\bfW'$ are generated in parallel). The updated value $\alpha_3^{(t+1)}$ is then projected onto an interval $[0,A_t]$ to guarantee the positivity constraint $\alpha_3 \in \mathbb{R}^+$ and the stability of the stochastic optimisation algorithm (we have used $A_t = 20$).

It is worth mentioning that if it was possible to simulate the auxiliary variables $(\bfS',\bfW')$ exactly from \eqref{eq:GMRF} then the estimator of $\nabla \log f(\MATpix | \alpha_3^{(t)})$ used in Algo. \ref{algo:algo1} would be unbiased and as a result $\alpha_3^{(t)}$ would converge exactly to $\hat{\alpha}_3$. However, exact simulation from \eqref{eq:GMRF} is not computationally feasible and therefore we resort to the MCMC kernel $\mathcal{K}$ and obtain a biased estimator of $\nabla \log f(\MATpix | \alpha_3^{(t)})$ that drives $\alpha_3^{(t)}$ to a neighbourhood of $\hat{\alpha}_3$ \cite{Pereyra2014ssp}. We have found that computing this biased estimator is significantly less expensive than alternative approaches, (e.g., using an approximate Bayesian computation algorithm as in \cite{Pereyra2013ip}), and that it leads to very accurate nonlinear unmixing results.

\section{Simulations: Synthetic data}
\label{sec:simu_synth}
In this section we study the performance of the proposed algorithm on a series of synthetic hyperspectral images firstly with linear mixing and secondly with nonlinear mixing. 
\subsection{First scenario: Linearly mixed image}
The objective here is to assess whether using the nonlinear unmixing model proposed in this paper leads to good unmixing results when analysing linearly mixed images, or if the additional degrees of freedom in the model degrade the estimation performance. This is crucial because in real hyperspectral images most pixels exhibit predominantly linear mixing.
We evaluate the performance of the proposed G-RCA algorithm (and its version incorporating the nonlinearity positivity constraints, G-RCA+) by unmixing a synthetic image of $100 \times 100$ pixels generated with the classical linear mixing model (i.e., (1) with $\bgam_{i,j} = 0$) and using $R=3$ endmembers (i.e., green grass, olive green paint and galvanised steel metal)\footnote{we extracted these endmembers from the spectral libraries of the ENVI software \cite{ENVImanual2003} in a similar fashion to previous work \cite{Eches2010,Halimi2010,Altmann2012a}}. This image is generated using $L=207$ spectral bands uniformly sampled from $400$nm to $2\,500$nm and with an average signal to noise ratio of $30$dB ($\sigma_{\ell}^2=3.10^{-4}, \forall \ell$). The abundance vectors $\Vabond{i,j}$ used to produce this image have been generated using the model \eqref{eq:abond_prior} (we later present another experiment where the abundances satisfy the sum-to-one constraint). The G-RCA and G-RCA+ algorithms for this experiment were implemented with $N_{\textrm{MC}}=2\,000$, $N_{\textrm{bi}}=1\,500$.  

The performance unmixing algorithms in terms of abundance estimation is evaluated by computing the root normalised mean square error (RNMSE) defined by
\begin{eqnarray}
\textrm{RNMSE}=\sqrt{ \dfrac{1}{N_{\textrm{row}} N_{\textrm{col}} R} \sum_{i,j} \norm{\Vabond{i,j}-\widehat{\boldsymbol a}_{i,j}}^2}
\end{eqnarray}
where $\Vabond{i,j}$ and $\widehat{\boldsymbol a}_{i,j}$ are the true and estimated abundance vectors for the pixel $(i,j)$ of the image.

For this scenario, the proposed G-RCA algorithm is compared with the classical NCLS algorithm \cite{Heinz2001} assuming the LMM (without sum-to-one constraint (STO)), comparisons to nonlinear SU methods will be addressed in scenario 2 described below. The results obtained with G-RCA,G-RCA+ and NCLS are $1.04\times10^{-2},1.04\times10^{-2}$ and $0.97\times10^{-2}$ respectively. We observe that the three methods performed similarly, showing that using G-RCA+ to analyse linearly mixed pixels does not degrade significantly the estimation performance. 

By analysing the distribution of the estimated nonlinearity levels $\widehat{\textbf s}_{i,j}^2$ (computed by approximating the expectation $\textrm{E}\left[\textbf{s}_{i,j}^2 | \MATpix,\hat{\alpha}_3\right]$) we confirm that G-RCA/G-RCA+ correctly identifies linearly mixed pixels. Indeed, the mean and variance of the estimated nonlinearity levels (computed by polling the $10000$ pixels) are $1.4 \times 10^{-4}$ and $1.9 \times 10^{-7}$ for G-RCA and $2.0 \times 10^{-5}$ and $4.2 \times 10^{-9}$ for G-RCA+, confirming that the amplitude of the nonlinear coefficients are significantly smaller than that of the abundances. It is also worth mentioning that unlike NCLS, G-RCA/G-RCA+ is able to handle unknown coloured noise (i.e., frequency-dependent noise levels).

\subsection{Second scenario: Nonlinear mixtures}
\begin{figure}[!ht]
\centering
\includegraphics[width=\columnwidth]{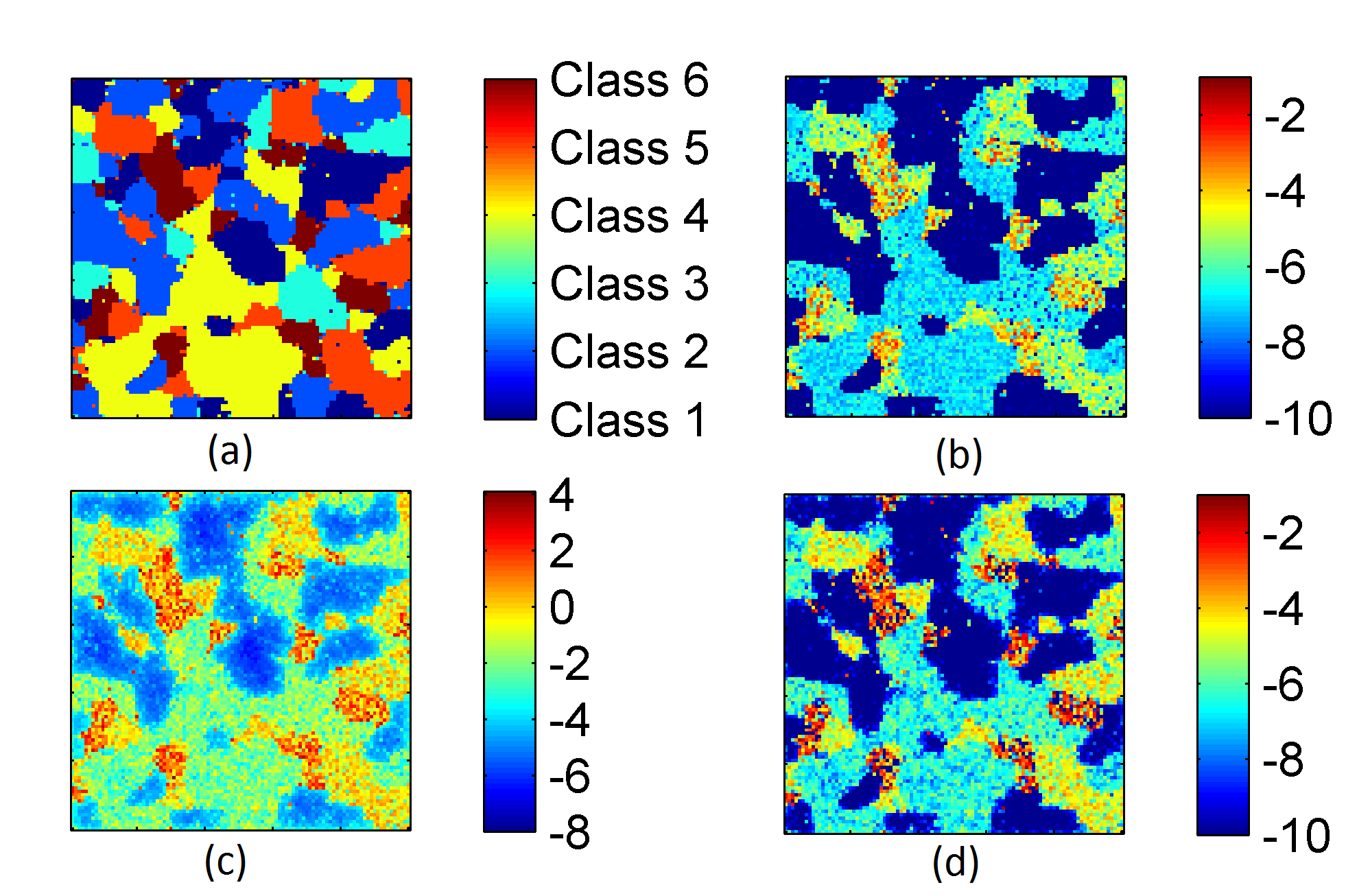}
 \caption{Nonlinear unximing: (a) Mixing model (class) allocation, (b) true log-energy of the nonlinear effects, (c)-(d) nonlinear log-energy estimated with G-RCA and with G-RCA+.}
 \label{fig:nonlin_synth}
\end{figure}

\underline{Data Set:}\\
The objective here is to evaluate the performance of the proposed model when applied to images containing different kinds of linear and nonlinear mixtures. We consider a synthetic image of $100 \times 100$ pixels generated with the same $R=3$ endmembers of the previous experiment, but using $6$ different mixing models. More precisely, we have used a Potts-Markov random field (with parameter $\beta = 1.6$) to generate a spatially coherent partition of the image where each partition is assigned to one of the $6$ mixing models, which was then used to generate the observations for that partition of the image, (the map with the mixing model assigned to each pixel is depicted in Fig. \ref{fig:nonlin_synth} (a)). The class $\mathcal{C}_1$ (resp. $\mathcal{C}_2$) is associated the LMM without (resp. with) abundance STO (LMM-WSTO and LMM-STO, respectively).
The pixels of class
$\mathcal{C}_3$ 
have been generated according to the
generalized bilinear mixing model (GBM) \cite{Halimi2010}
\begin{eqnarray}
\Vpix{i,j}  = \sum_{r=1}^{R} \abond{r}{i,j}\Vmat{r} + \sum_{k=1}^{R-1} \sum_{k'=k+1}^{R}\gamma_{i,j}^{(k,k')}\abond{k}{i,j} \abond{k'}{i,j}\Vmat{k}\odot\Vmat{k'}  + \Vnoise_{i,j}
\end{eqnarray}
with $\gamma_{i,j}^{(k,k')}=1$, which corresponds to the model investigated in \cite{Fan2009} (Fan's model).
The class $\mathcal{C}_4$ is composed of pixels
generated according to the PPNMM \cite{Altmann2012a} as follows
\begin{eqnarray}
\Vpix{i,j} & = & \MATmat \Vabond{i,j} + b \left(\MATmat \Vabond{i,j} \right) \odot \left(\MATmat \Vabond{i,j} \right) + \Vnoise_{i,j}. 
\end{eqnarray}
with $b=0.2$.
The pixels of the class $\mathcal{C}_5$ have been generated using the bilinear model investigated in \cite{Nascimento2009}, referred to as Nascimento's model (NM) and defined as 
\begin{eqnarray}
\label{eq:NM}
\Vpix{i,j}  =  \sum_{r=1}^{R} \abond{r}{i,j}\Vmat{r} + \sum_{k=1}^{R-1} \sum_{k'=k+1}^{R}\gamma_{i,j}^{(k,k')}\Vmat{k}\odot\Vmat{k'}  + \Vnoise_{i,j}
\end{eqnarray}
where the mixture coefficients (associated with the linear and nonlinear terms) of each pixel sum to one.
Finally, the class $\mathcal{C}_6$ has been generated according to \eqref{eq:NLM0} with zero-mean Gaussian nonlinearity coefficients with variance $s_{i,j}^2=0.1$. Note that $\mathcal{C}_6$ is the only class allowing for negative nonlinearities. For the classes $\mathcal{C}_2$, $\mathcal{C}_3$ and $\mathcal{C}_4$ (whose underlying models rely on the abundance STO assumption), the abundance vectors have
been randomly generated according to a uniform distribution
over the admissible set defined by the positivity and sum-to-
one constraints. The mixing coefficients in \eqref{eq:NM} (for the pixels of $\mathcal{C}_5$) have been uniformly generated in the simplex defined by the positivity and STO constraints. The abundances of the pixels in $\mathcal{C}_1$ and $\mathcal{C}_6$ have been generated according to \eqref{eq:abond_prior} with $\beta_r=0.3, \forall r$.
All pixels have been corrupted by additive
i.i.d Gaussian noise of variance $\sigma^2=3 \times 10^{-4}$, corresponding to
an average signal-to-noise ratio SNR
$29$dB. The noise is
assumed to be i.i.d. for a fair comparison with SU algorithms
assuming i.i.d. Gaussian noise. Fig. \ref{fig:nonlin_synth} (b) shows the log-energy
of the nonlinear contribution for each pixel of the image, i.e., $\log(\bphi_{i,j})$. 

\underline{Unmixing:}

\begin{table*}[ht]
\renewcommand{\arraystretch}{1.2}
\begin{footnotesize}
\begin{center}
\begin{tabular}{|c|c|c|c|c|c|c|c|}
\cline{3-8}
\multicolumn{2}{c|}{}& $\mathcal{C}_1$ & $\mathcal{C}_2$ & $\mathcal{C}_3$ & $\mathcal{C}_4$ & $\mathcal{C}_5$ & $\mathcal{C}_6$\\
\multicolumn{2}{c|}{}& (LMM-WSTO) & (LMM-STO) & (GBM) & (PPNM) & (NM) & (RCA)\\
\hline
\multirow{6}{*}{SU Algo.}& NLCS & $\red{\textbf{0.98}}$ & $0.96$ & $5.11$ & $5.10$ & $10.38$ & $26.35$\\
\cline{2-8}
& FLCS & $81.45$ & $\red{\textbf{0.59}}$ & $11.57$ & $9.90$ & $32.51$ & $30.40$\\
\cline{2-8}
& GBM & $80.68$ & $\blue{\textbf{0.60}}$ & $4.64$ & $5.01$ & $32.54$ & $29.25$\\
\cline{2-8}
& PPNM & $70.33$ & $1.11$ & $\red{\textbf{1.85}}$ & $\red{\textbf{0.97}}$ & $28.13$ & $23.08$\\
\cline{2-8}
& NM & $81.06$ & $0.98$ & $11.53$ & $9.77$ & $\red{\textbf{2.73}}$ & $29.43$\\
\cline{2-8}
& RCA & $\blue{\textbf{1.12}}$ & $1.09$ & $2.69$ & $\blue{\textbf{2.62}}$ & $3.63$ & $\red{\textbf{6.85}}$\\
\cline{2-8}
& G-RCA & $1.34$ & $1.29$ & $2.69$ & $2.65$ & $3.56$ & $\blue{\textbf{6.96}}$\\
\cline{2-8}
& G-RCA+& $1.21$ & $1.14$ & $\blue{\textbf{2.11}}$ & $2.86$ & $\blue{\textbf{2.88}}$ & $19.63$ \\
\hline
\end{tabular}
\vspace{0.2cm}
\caption{Scenario 2: Abundance RNMSEs ($\times 10^{-2}$).\label{tab:RNMSE_synth}}
\end{center}
\end{footnotesize}
\end{table*}

\begin{table*}[ht]
\renewcommand{\arraystretch}{1.2}
\begin{footnotesize}
\begin{center}
\begin{tabular}{|c|c|c|c|c|c|c|c|}
\cline{3-8}
\multicolumn{2}{c|}{}& $\mathcal{C}_1$ & $\mathcal{C}_2$ & $\mathcal{C}_3$ & $\mathcal{C}_4$ & $\mathcal{C}_5$ & $\mathcal{C}_6$\\
\multicolumn{2}{c|}{}& (LMM-WSTO) & (LMM-STO) & (GBM) & (PPNM) & (NM) & (RCA)\\
\hline
\multirow{6}{*}{SU Algo.}& NLCS & $\red{\textbf{1.72}}$ & $\blue{\textbf{1.72}}$ & $1.81$ & $1.78$ & $2.07$ & $4.43$\\
\cline{2-8}
& FLCS & $55.12$ & $\blue{\textbf{1.72}}$ & $2.61$ & $2.49$ & $7.65$ & $11.69$\\
\cline{2-8}
& GBM & $54.54$ & $\blue{\textbf{1.72}}$ & $1.92$ & $2.01$ & $7.65$ & $11.28$\\
\cline{2-8}
& PPNM & $\blue{\textbf{9.47}}$ & $\blue{\textbf{1.72}}$ & $\blue{\textbf{1.73}}$ & $\blue{\textbf{1.72}}$ & $3.26$ & $4.00$\\
\cline{2-8}
& NM & $55.09$ & $1.73$ & $2.62$ & $2.62$ & $\blue{\textbf{1.71}}$ & $9.97$\\
\cline{2-8}
& RCA & $\red{\textbf{1.72}}$ & $\red{\textbf{1.71}}$ & $\red{\textbf{1.70}}$ & $\red{\textbf{1.71}}$ & $\red{\textbf{1.70}}$ & $\red{\textbf{1.70}}$\\
\cline{2-8}
& G-RCA & $\red{\textbf{1.72}}$ & $\red{\textbf{1.71}}$ & $\red{\textbf{1.70}}$ & $\red{\textbf{1.71}}$ & $\red{\textbf{1.70}}$ & $\red{\textbf{1.70}}$\\
\cline{2-8}
& G-RCA+ &$\red{\textbf{1.72}}$ & $\blue{\textbf{1.72}}$ & $\red{\textbf{1.72}}$ & $\blue{\textbf{1.72}}$ & $\blue{\textbf{1.72}}$ & $\blue{\textbf{3.60}}$\\
\hline
\end{tabular}
\vspace{0.2cm}
\caption{Scenario 2: Reconstruction errors ($\times 10^{-2}$).\label{tab:RE_synth}}
\end{center}
\end{footnotesize}
\vspace{-0.4cm}
\end{table*}
Different estimation procedures have been
considered for the four different mixing models:
\begin{itemize}
\item The NCLS algorithm \cite{Heinz2001} which is known to have good
performance for linear mixtures when the abundance STO assumption can be relaxed.
\item The FCLS algorithm \cite{Heinz2001} which is known to have good
performance for linear mixtures and relying on the abundance STO.
\item The GBM-based optimization approach \cite{Halimi2010} which is
adapted for bilinear nonlinearities. The optimization algorithm is stopped when the norm of the difference between consecutive parameter estimates is smaller than $10^{-6}$.
\item The gradient-based approach of \cite{Altmann2012a} which is based on
a PPNMM and has shown nice properties for various polynomial nonlinearities.
\item The FCLS algorithm used with an extended endmember matrix (containing the bilinear products of the endmembers) for unmixing based on the NM. This algorithm is denoted by NM in the remainder of the paper.
\item The RCA algorithm proposed in \cite{Altmann2014a} with $K=6$ classes, $N_{\textrm{MC}}=3000$, $N_{\textrm{bi}}=200$ and $\beta=1.6$ ($\beta$ is the granularity parameter of the Potts model used in \cite{Altmann2014a} for nonlinearity-based segmentation).
\item The proposed G-RCA/G-RCA+ with $N_{\textrm{MC}}=2000$, $N_{\textrm{bi}}=1500$. 
\end{itemize}
The abundance estimation performance of $6$ unmixing strategies is evaluated using the RNMSE obtained for each class of pixels and defined by 
\begin{eqnarray}
\textrm{RNMSE}_k=\sqrt{ \dfrac{1}{N_k R} \sum_{(i,j) \in I_k} \norm{\Vabond{i,j}-\widehat{\boldsymbol a}_{i,j}}^2}
\end{eqnarray}
where $I_k$ is the set of indices of the pixels in the class $\mathcal{C}_k$ and $N_k$ is the number of pixels in $\mathcal{C}_k$. The results obtained by the $6$ algorithms and are presented in Table \ref{tab:RNMSE_synth} which shows that the model \eqref{eq:NLM0}-\eqref{eq:nonlin} leads to robust abundance estimation algorithms since all RCA, G-RCA and G-RCA+ provide satisfactory results for the 6 different mixtures. Moreover, we observe that G-RCA+ and G-RCA perform almost as well as RCA; note that G-RCA+ and G-RCA are fully unsupervised, whereas RCA requires specification of the parameters $K$ and $\beta$ (note that the results in Table \ref{tab:RNMSE_synth} have been obtained with the best values for $K$ and $\beta$ identified by cross-validation, using different values could have degraded estimation performance significantly). We also observe that G-RCA+ generally provides better abundance estimates than G-RCA, supporting the assumption that $\gamma_{i,j} \in \mathbb{R}^+$. Finally, Fig. \ref{fig:nonlin_synth}, (c) and (d) show the nonlinearity level map estimated by G-RCA and G-RCA+ (computed from the estimates of $\{\norm{\bphi_{i,j}}^2 \}$). We observe that both G-RCA+ and G-RCA were able to correctly identify the regions where nonlinear effects occur, and that G-RCA+ produced a better estimate of the nonlinearity energies.

In order to assess the capacity of GRCA+ and GRCA to fits different types of mixing models, Table \ref{tab:RE_synth} reports the average image reconstruction (RE) for each class of pixels and defined by 
\begin{eqnarray}\label{recError}
\textrm{RE}_k=\sqrt{ \dfrac{1}{N_k L} \sum_{(i,j) \in I_k} \norm{\Vpix{i,j}-\widehat{\mathbf y}_{i,j}}^2}
\end{eqnarray}
where $\widehat{\mathbf y}_{i,j}$ is the $(i,j)$th reconstructed pixel. 
The reconstruction errors in Table \ref{tab:RE_synth} confirm that RCA and G-RCA are very flexible and can handle the $6$ different mixing models associated with the $6$ classes considered. More precisely, the reconstruction errors provided by RCA and G-RCA correspond to the standard deviation of the additive Gaussian noise ($\sigma^2=3 \times 10^{-4}$). G-RCA+ also provides accurate reconstructions errors, except for class $\mathcal{C}_6$, because G-RCA+ does not consider negative nonlinearities. 

 \begin{figure*}
\centering
\includegraphics[width=\columnwidth]{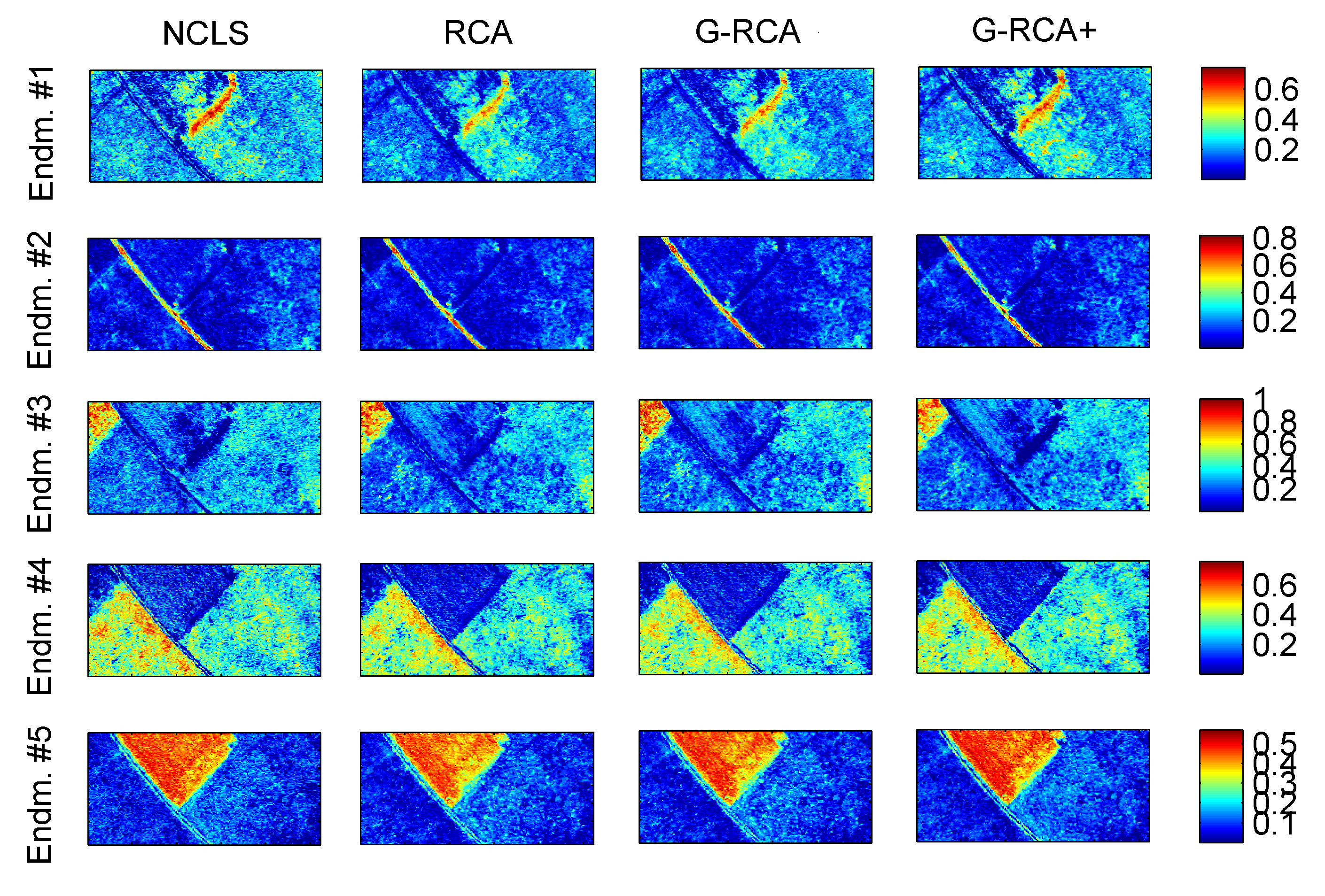}
 \caption{Abundance maps estimated with NCLS, RCA, G-RCA and G-RCA+ (from left to right) for the Villelongue real image.}
 \label{fig:abund}
\end{figure*}

\begin{figure}[!ht]
\centering
\includegraphics[width=\columnwidth]{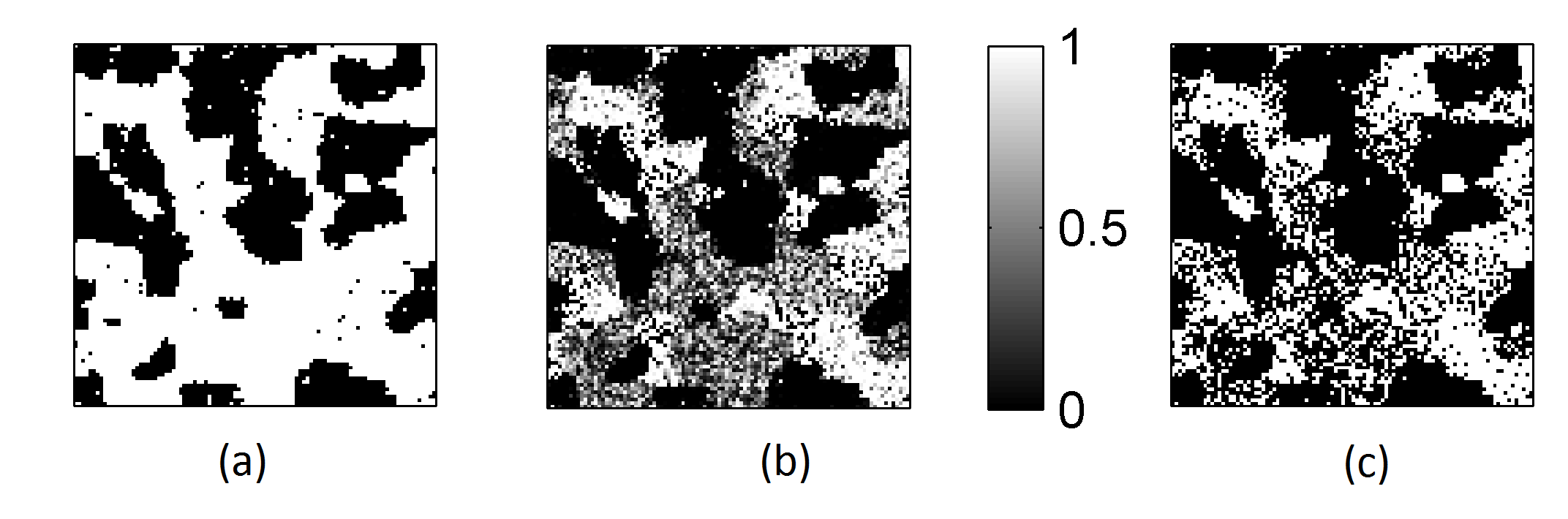}
 \caption{Nonlinear mixing detection in the synthetic image of Scenario 2: (a) Nonlinearity presence/absence (ground truth), (b) Posterior probability of significant nonlinearities $P_{i,j}$ estimated with G-RCA+ using \eqref{eq:nonlin_proba}, (c) Bayesian hypothesis test for nonlinearity detection ($P_{i,j} > 1/2$).}
 \label{fig:detection_scenar2}
\end{figure}

\underline{Bayesian detection of nonlinearity}\\
As explained previously, the proposed MCMC algorithm can be used to detect image pixels with significant nonlinear mixing (this is formulated as a Bayesian hypothesis test involving the posterior probability \eqref{eq:nonlin_proba}). In order to illustrate this, Fig. \ref{fig:detection_scenar2} compares the true nonlinearity presence map (depicted in Fig. \ref{fig:detection_scenar2} (a)) with the probabilities \eqref{eq:estim_nonlin_proba} estimated with G-RCA+ and $\eta=2$ (Fig. \ref{fig:detection_scenar2} (b)) and the corresponding detection map (Fig. \ref{fig:detection_scenar2} (c)) for the synthetic image considered in Scenario 2. Recall that the detection map is computed by thresholding the probability map (we used $a_0 = a_1$ leading to a threshold value of $0.5$). It is worth noting that both $\eta$ in \eqref{eq:nonlin_proba} and the loss function coefficients $a_0, a_1$ are application specific and can be adjusted to reflect prior knowledge about the confidence in the model for a specific scene, the probability of nonlinear effects, and the relative cost of false positive (false alarm) and false negative detections. Moreover, Table \ref{tab:detection} shows the the empirical probability of false alarm $P_{\textrm{FA}}$ and probability of detection $P_{\textrm{D}}$ computed with G-RCA+ for this experiment and using different values of $\eta$. We observe the good performance of G-RCA+, which for $\eta = 1$ is able to detect over $85\%$ of nonlinearly mixed pixels with a probability of false alarm of $0.5\%$. Finally, it is of note that unlike the original RCA algorithm \cite{Altmann2014a}, the Bayesian model proposed in this paper does not allocate prior probability to the specific case of linearly mixed pixels (i.e., the case $\bphi_{i,j}=\Vzero$ has prior density but not prior mass). As a result, it is not possible to perform Bayesian point hypothesis tests (i.e., $\bphi_{i,j}=\Vzero$ vs $\bphi_{i,j}\neq\Vzero$) that are possible with RCA. The development of a new Bayesian model and Bayesian tests that combine the strengths of both approaches is  currently under investigation.

\begin{table}[ht]
\renewcommand{\arraystretch}{1.2}
\begin{footnotesize}
\begin{center}
\begin{tabular}{|c|c|c|c|c|c|}
\cline{2-6}
\multicolumn{1}{c|}{}& $\eta=1$ & $\eta=1.5$ & $\eta=2$ & $\eta=2.5$ & $\eta=3$\\
\hline
$P_{\textrm{FA}} (\times10^{-2})$ & $0.53$ & $0.15$ & $0.30$ & $0.02$ & $0$\\
\hline
$P_{\textrm{D}} (\times10^{-2})$   & $85.83$ & $78.93$ & $72.33$ & $66.15$ & $60.95$\\
\hline
\end{tabular}
\vspace{0.2cm}
\caption{Scenario 2: Detection performance.\label{tab:detection}}
\end{center}
\end{footnotesize}
\vspace{-0.4cm}
\end{table}

\section{Simulations: Real hyperspectral image}
\label{sec:simu_real}

\begin{figure}[!ht]
\centering
\includegraphics[width=\columnwidth]{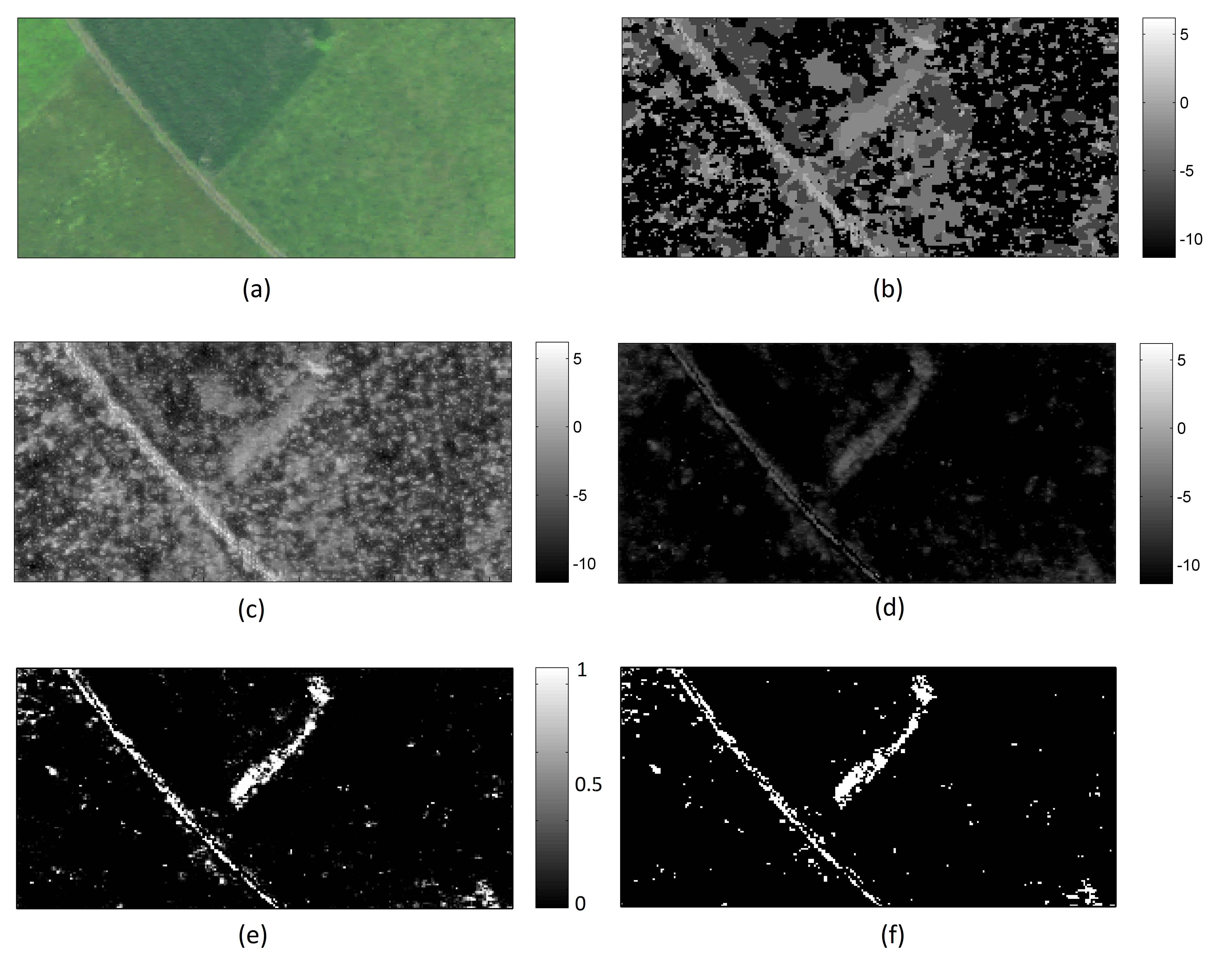}
 \caption{Nonlinearity level estimation: (a) True colour image of the scene of interest. Levels of nonlinearity estimated with RCA (using $K=5$ levels) (b), with G-RCA (c) and with G-RCA+ (d). (e) G-RCA+ estimation of posterior probability of significant nonlinear mixing \eqref{eq:nonlin_proba} ($\eta = 2$), and (f) nonlinear mixing detection map.}
 \label{fig:detection_Madonna}
\end{figure}

This section presents an application of the proposed  G-RCA method to a real hyperspectral image. The hyperspectral image considered in this experiment was acquired in 2010 by the Hyspex hyperspectral scanner over Villelongue, France ($00^\circ03$'W and $42^\circ57$'N). This scene was observed at $L = 160$ spectral bands
ranging from the visible to near infra-red with a spatial resolution of $0.5$m. This dataset has already been studied in \cite{Sheeren2011,Altmann2013,Altmann2014a,Altmann2014b} and is mainly composed of forested and urban areas (see \cite{Sheeren2011} for more details about the data acquisition and
pre-processing steps). We have applied our method to the region of interest of size $180 \times 250$ pixels that is depicted in Fig.
\ref{fig:detection_Madonna} (a). This region is composed mainly of a path and different vegetation species and has $R=5$ endmembers, whose spectral signatures have been extracted from the data using VCA \cite{Nascimento2005}. 

Figs. \ref{fig:detection_Madonna} (c) and (d) show the nonlinearity levels estimated with the proposed G-RCA/G-RCA+ method. For comparison, the results obtained with RCA \cite{Altmann2014a} are presented in Fig. \ref{fig:detection_Madonna} (d) (recall that to simplify the estimation problem, RCA artificially constrains nonlinearities to take a finite number of values (5 here)). Since RCA does not directly estimate $\{\norm{\bphi_{i,j}}^2 \}$ but $\{\norm{s_{i,j}}^2 \}$, Figs. \ref{fig:detection_Madonna} (c) and (d) depicts the minimum mean square error estimator of the pixel-wise nonlinearity level 
\begin{eqnarray}\label{S_MMSE}
\hat{\bfS}_{\textrm{MMSE}} = \textrm{E}\left[\bfS| \MATpix,\hat{\alpha}_3\right],
\end{eqnarray}
where the expectation is now taken with respect to the marginal posterior density $f(\bfS |\MATpix,\alpha_3)$. In a similar fashion to the abundance estimators, these estimators are approximated using Monte Carlo using
\begin{eqnarray}
\label{eq:S_MC}
\hat{\bfS}_{MMSEj} = \dfrac{1}{N_{\textrm{MC}}-N_{\textrm{bi}}}\sum_{t=N_{\textrm{bi}}+1}^{N_{\textrm{MC}}}
	\bfS^{(t)}.
\end{eqnarray}
We observe that the results obtained with both methods are in good visual agreement and highlight spatial structures that can be easily identified in the colour image (e.g., path) where one would expect nonlinear mixing to occur. More importantly, by not constraining the number of nonlinearity levels, G-RCA and G-RCA+ produce spatially smooth estimates that are realistic (and that do not require specification of the number of nonlinearity levels a priori). It is important to note that the results obtained with G-RCA+ indicate that the nonlinear effects in the image are sparser and weaker than previously suggested by RCA (and G-RCA). Due to the high correlation between the nonlinear terms (cross-products of the endmembers) as well as their energy, the estimation of the nonlinear coefficients is difficult, particularly for RCA and G-RCA that do no take into account the positivity of $\bGam$. Indeed, in the model used in RCA and G-RCA $\bGam$ can take large positive and negative values which average out, leading to large estimated nonlinearity levels. By constraining the nonlinear coefficients to be non-negative, G-RCA+ yields smaller and sparser nonlinearity levels that are arguably closer to the ground truth. Finally, Figs. \ref{fig:detection_Madonna} (e)-(f) show the estimated posterior probability of nonlinear mixing computed using  \eqref{eq:nonlin_proba} (with $\eta=2$) and the nonlinear mixing detection results (computed by thresholding the probabilities w.r.t. $0.5$). Again, nonlinearly mixed pixels are clearly identified near the path and at the boundary between two fields where we expect nonlinear mixing to occur.

The abundance maps obtained by G-RCA/G-RCA+ have been compared to those obtained with the algorithm considered in Section V and the results obtained by the different methods are generally similar. As an example, Fig. \ref{fig:abund} shows that the abundances estimates obtained with RCA, G-RCA, G-RCA+ (RCA-based nonlinear models) and with the NCLS (LMM-based) algorithm. Note that there is no abundance ground truth for this image making it difficult to quantify abundance estimation precision directly. This figure shows that algorithm based on the RCA model provide abundance maps generally in agreement with those obtained with NCLS, although the results can vary locally (e.g, abundances of the first endmember between the two fields). We implemented G-RCA(+) using $N_{\textrm{MC}}=2000$ and $N_{\textrm{bi}}=1500$ (computing these results using MATLAB required $7$ hours on a $3$GHz Intel Xeon quad-core workstation). We observe that G-RCA and G-RCA+ perform similarly to the RCA algorithm, while not requiring fixing a number of classes. Finally, for this image all the methods achieved the same reconstruction error $\textrm{RE}= \sqrt{(\sum_{i,j} {\norm{\hat{\Vpixels}_{i,j} - \Vpix{i,j}}^2})/(N_{\textrm{row}}N_{\textrm{col}}L)} = 0.22$, due to the fact that the image is predominantly composed of linearly mixed pixels for which the eight methods perform similarly.



\section{Conclusion}
\label{sec:conclusion}
This paper has presented a new hierarchical Bayesian algorithm for spectral unmixing of 
hyperspectral images which incorporates the spatial dependencies inherent in an image associated with the nonlinear mixture effects.
The nonlinear mixtures were decomposed into a linear combination of the endmembers 
and an additive term which represents the nonlinear effects. This term was further decomposed 
as a combination of the endmembers cross-products. A Gamma Markov random field was 
introduced to promote smooth nonlinearity variations in the image. In contrast with previously reported 
work where nonlinear unmixing relied on a nonlinearity level-based image segmentation, 
the proposed model allows the level of nonlinearity to differ in each pixel while 
allowing the identification of regions where nonlinear effects occur. In this paper, 
a zero-mean Gaussian prior, restricted to the positive orthant was assigned to the nonlinear coefficients 
of each pixel. This choice was motivated by the fact that several existing models include positivity constraints for the nonlinear terms, e.g. 
\cite{Somers2009,Nascimento2009,Halimi2010}, include such constraints within the 
SU procedure, and this was previously not possible using the RCA model in \cite{Altmann2014a} due to the marginalisation of these parameters. The results presented in this paper have shown that it can significantly improve the unmixing performance. In this paper, the endmembers were assumed to be perfectly known but often need to be extracted from the data. Future work will include the generalisation of the G-RCA+ model to account for endmember estimation errors and more general sources of nonlinearity (such as endmember intrinsic variability). Finally, in some images abundances exhibit strong spatial correlations, and taking this information into account may improve estimation performance significantly \cite{Eches2011,Iordache2012,Chen2014}. Therefore it would be very interesting to design new models and nonlinear unmixing procedures that are capable of simultaneously exploiting the spatial correlations between abundances and between nonlinearities to produce best results.
\bibliographystyle{IEEEbib}
\bibliography{biblio1}

\end{document}